\documentclass[a4paper,twocolumn,11pt,accepted=2021-10-18]{quantumarticle}
\pdfoutput=1

\usepackage[utf8]{inputenc}
\usepackage[english]{babel}
\usepackage[T1]{fontenc}
\usepackage[numbers,sort&compress]{natbib}
\usepackage{lipsum}

\usepackage{authblk}

\usepackage[margin=1in]{geometry}

\usepackage{enumerate}

\usepackage{microtype}

\usepackage{tikz}
\usetikzlibrary{backgrounds,fit,decorations.pathreplacing,calc}

\usepackage{soul}

\usepackage{float}

\usepackage[colorlinks = true]{hyperref}
\definecolor{darkred}  {rgb}{0.5,0,0}
\definecolor{darkblue} {rgb}{0,0,0.5}
\definecolor{darkgreen}{rgb}{0,0.5,0}
\hypersetup{
  urlcolor   = blue,         
  linkcolor  = darkblue,     
  citecolor  = darkgreen,    
  filecolor  = darkred       
}

\usepackage{amsmath,amssymb,amsfonts,amsthm,amstext}
\usepackage{bbm}

\usepackage{etoolbox}

\usepackage{mathtools}
\mathtoolsset{centercolon}
\makeatletter
\protected\def\tikz@nonactivecolon{\ifmmode\mathrel{\mathop\ordinarycolon}\else:\fi}
\makeatother

\usepackage{tcolorbox}

\usepackage{algorithm}
\usepackage{algpseudocode}

\usepackage{cleveref}

\crefname{lemma}{Lemma}{Lemmas}
\crefname{proposition}{Proposition}{Propositions}
\crefname{definition}{Definition}{Definitions}
\crefname{theorem}{Theorem}{Theorems}
\crefname{conjecture}{Conjecture}{Conjectures}
\crefname{corollary}{Corollary}{Corollaries}
\crefname{claim}{Claim}{Claims}
\crefname{section}{Section}{Sections}
\crefname{appendix}{Appendix}{Appendices}
\crefname{figure}{Fig.}{Figs.}
\crefname{table}{Table}{Tables}

\usepackage[retainorgcmds]{IEEEtrantools}

\usepackage{tocloft}

\usepackage{multirow}

\usepackage{tikz}	
\usetikzlibrary{backgrounds,fit,decorations.pathreplacing}

\newcommand{\ket}[1]{|#1\rangle}
\newcommand{\bra}[1]{\langle#1|}
\newcommand{\braket}[2]{\langle#1|#2\rangle}
\newcommand{\ketbra}[2]{|#1\rangle\langle#2|}

\newcommand{\x}{\otimes}

\newcommand{\ct}{^{\dagger}}

\DeclarePairedDelimiter{\set}{\lbrace}{\rbrace}
\DeclarePairedDelimiter{\abs}{\lvert}{\rvert}
\DeclarePairedDelimiter{\norm}{\lVert}{\rVert}

\DeclarePairedDelimiter{\ip}{\langle}{\rangle}

\DeclareMathOperator{\spn}{span}

\newcommand{\C}{\mathbb{C}}
\newcommand{\R}{\mathbb{R}}

\newcommand{\Z}{\mathbb{Z}}
\newcommand{\calH}{\mathcal{H}}

\newcommand{\calU}{\mathcal{U}}

\newcommand{\Zp}{\Z_p^{\ast}}

\newcommand{\1}{\mathbb{I}}

\newcommand{\CHSH}{\ensuremath{\mathsf{CHSH}}}
\newcommand{\MIP}{\ensuremath{\mathsf{MIP}}}

\newcommand{\tP}{\tilde{P}}
\newcommand{\tQ}{\tilde{Q}}

\newcommand{\tpsi}{\tilde{\psi}}

\newcommand{\nr}{n(r)}
\newcommand{\mr}{m(r)}

\newcommand{\pr}[2]{P(#1|#2)}

\newcommand{\ep}{\epsilon}

\newtheorem{theorem}{Theorem}[section]
\newtheorem{theorem*}{Theorem}
\newtheorem{lemma}[theorem]{Lemma}
\newtheorem{proposition}[theorem]{Proposition}
\newtheorem{definition}[theorem]{Definition}

\newtheorem*{conjecture*}{Conjecture}

\theoremstyle{definition}

\begin{document}

\title{Constant-sized correlations are sufficient to self-test maximally entangled states with unbounded dimension}

\author{Honghao Fu}
\email{h7fu@umd.edu}
\affiliation{Joint Center for Quantum Information and Computer Science,
Institute for Advanced Computer Studies and Department of Computer Science, 
University of Maryland, College Park, MD 20742, USA}
\maketitle

\begin{abstract}
    Let $p$ be an odd prime and let $r$ be the smallest generator of the multiplicative group $\Zp$.
	We show that there exists a correlation of size $\Theta(r^2)$ 
	that self-tests a maximally entangled state of local dimension $p-1$.
	The construction of the correlation uses the embedding procedure
	proposed by Slofstra (\textit{Forum of Mathematics, Pi.} ($2019$)).
	Since there are 
	infinitely many prime numbers whose smallest multiplicative generator
	is in the set $\set{2,3,5}$ (
	D.R. Heath-Brown \textit{The Quarterly Journal of Mathematics}
	($1986$) and
	M. Murty \textit{The Mathematical Intelligencer} ($1988$)), 
	our result implies that constant-sized correlations are sufficient for self-testing of maximally entangled states
	with unbounded local dimension.
\end{abstract}

The certification of a quantum device is an important building block
for many quantum information processing tasks, especially
when the devices are provided by some untrusted vendor.
We would like such certification to be done based solely on
the observed measurement statistics and with the only assumption that
any local device cannot communicate with the other local devices. 
The measurement statistics are referred to as correlations.
It has been shown that certain quantum correlations require the distant parties to share
a particular entangled state up to some local isometry. 
This phenomenon is referred to as self-testing.  

The fact that the verifier only interacts classically with the unknown device
makes self-testing a powerful tool 
for applications in 
quantum cryptography and computational
complexity theory.
It allows a classical party to delegate quantum computations to some untrusted service provider
and verify that the computations are performed
honestly and correctly \cite{ruv2013,coladan2017verifier}.
Self-testing also becomes a critical component of the security proofs of device-independent quantum cryptographic protocols
\cite{mayersyao,fu2018}.
Since the setting of multi-prover interactive proof systems ($\MIP$) is similar to a self-test, self-tests also help to 
bound the computational power of entanglement-assisted
$\MIP$ protocols \cite{fitzsimons2019, neexp, re}.

The case of self-testing of the EPR pair
is fully understood.
The techniques for this case were 
first introduced in \cite{mckague2012}, then improved in \cite{bamps2015}. 
Self-testing of tensor products of maximally entangled qubits 
were proved in \cite{natarajan2017,lowdegree},
with the latter being the one with the smallest question and answer sets.
Self-testing of a general bipartite entangled states with local dimension $d$ was proved in \cite{coladan2017all},
which
uses four questions but each question had $d$ answers.
The number of questions was later reduced to two in \cite{sarkar2019},
but the number of answers was still $d$.

Our work aims at minimizing the sizes of the question 
and answer sets of a correlation that can self-test
a maximally entangled state with large local dimension.
For comparison, all the correlations used in the results listed above
have either the number of questions or answers dependent on the
local dimension of the entangled state.
Our progress is summarized in the following theorem.
\begin{theorem*}
\label{thm:inf}
    Let $r \in \set{2,3,5}$ and let $D$
    denote the infinite set of all odd primes $p$ whose smallest primitive root is $r$.
    Then for any $p \in D$, 
    the following maximally entangled state of local dimension $p-1$,
    \begin{align*}
        \ket{\tpsi} = \frac{1}{\sqrt{p-1}}
        \sum_{j=1}^{p-1} \ket{j}\ket{p-j},
    \end{align*}
	can be self-tested 
	with constant-sized question and answer sets.
\end{theorem*}

\subsection*{Proof overview}

To prove \Cref{thm:inf}, we construct a bipartite quantum correlation $P_{p,r}$.
We denote the two parties by Alice and Bob.
Alice and Bob's question sets are of size $\Theta(r)$ and their answer sets are constant-sized
for each odd prime number $p$ with smallest primitive root $r$.
We say the size of the question set is of order $\Theta(r)$ to mean that 
there exist constants $c$ and $d$ such that
the size of the question set is $cr+d$, and
we say that $r$ is a primitive root of $p$ if $r$ is a 
multiplicative generator of the group $\Zp$.
Note that when $r \in \set{2,3,5}$ the size of neither the question set nor the answer set of $P_{p,r}$ depends on $p$. 
We can view $P_{p,r}$ in a matrix where each entry in the matrix is labelled by Alice's and Bob's question-answer pair, so the size
of $P_{p,r}$ is the size of this matrix, which equals the product of the number of Alice's question-answer pairs and the number of Bob's question answer pairs. Therefore, the size of
$P_{p,r}$ is of order $\Theta(r^2)$.
We prove $P_{p,r}$ self-tests a maximally entangled
state of dimension $(p-1)$.

The correlation $P_{p,r}$ contains a perfect correlation associated with a binary linear system.
To induce such a correlation, Alice is given a random equation 
from a linear system and she
should give an assignment to each variable of the equation.
Bob is asked to assign a value to a random variable of the chosen equation.
The correlation is perfect if Alice's assignment satisfies the equation and Bob's
assignment to the variable matches Alice's assignment.
A widely-used and thoroughly-studied example is the perfect correlation of 
the following linear system 
\begin{align*}
    &x_1 + x_2 + x_3 = 0 && x_4 + x_5 + x_6 = 0\\
    &x_7 + x_8 + x_9 = 0 
    &&x_1 + x_4 + x_7 = 0\\
    & x_2 + x_5 + x_8 = 1 &&
    x_3 + x_6 + x_9 = 0.
\end{align*}
The game version of this linear system is know as the Magic square game \cite{mermin1990,magic_square}.
Using two copies of $\ket{EPR}$, the perfect correlation can be induced.
It has been shown that if a strategy can induce this correlation, 
the shared state must be $\ket{EPR}^{\x 2}$ up to some local isometry \cite{wu2016}.
The key observation that leads to the self-testing proof is that 
an inducing strategy must contain binary observables $X$ and $Z$ that 
satisfy the anti-commutation relation 
\begin{align*}
    Z X Z = - X.
\end{align*}
Intuitively, we can think of this perfect correlation as a way to enforce the
anti-commutation relation \cite{coladan2017}.

We design a linear system $\hat{A}x=0$ such that the perfect correlation $P_{\hat{A}}$ associated with the
linear system can enforce the relation 
\begin{align}
\label{eq:uor}
    U O U\ct = O^r,
\end{align}
for unitaries $U$ and $O$.
The inspiration comes from Slofstra's seminal work \cite{slofstra2017},
where he outlined a new way 
to design binary linear system such that the associated perfect correlation can enforce conjugacy relations of the
form $X Y X\ct = Z$ for unitaries $X, Y$ and $Z$.
Following Slofstra's design, 
the numbers of equations and variables of $\hat{A}x=0$ are of order $\Theta(r)$.

The reason that we choose \Cref{eq:uor} to be the relation enforced
by $P_{\hat{A}}$ is the following.
Inducing $P_{\hat{A}}$ guarantees that the strategy contains unitary operators
$U$ and $O$ satisfying \Cref{eq:uor}.
Moreover, if we can certify that the unitary $O$
has the eigenvalue $\omega_p := e^{i 2\pi /p }$
, \Cref{eq:uor}
automatically guarantees that the spectrum of $O$ contains 
$\set{ \omega_p^j | 1 \leq j \leq p-1}$.
It also guarantees that the eigenspace of each distinct eigenvalue
has the same dimension. 

In order to certify an eigenvalue of $O$, we use the optimal correlation $P_{\mu}$
that achieves the maximal violation of the $\cot(\mu)$-weighted
$\CHSH$ inequality proposed in \cite{lawson2010, acin2012}
as a variation of the $\CHSH$ inequality \cite{chsh}
\begin{equation}
\label{eq:chsh}
\begin{aligned}
	I_{\cot(\mu)} =& \cot(\mu)\ip{M_0N_0 +M_0N_1} + \ip{M_1N_0-M_1N_1} \\
		 \leq &2\abs{\cot(\mu)},
\end{aligned}
\end{equation}
for binary observables $M_x, N_y$ with $x,y \in \set{0,1}$.
It has been proved that this correlation can self-test $\ket{EPR}$ \cite{jed2017}.
We further prove that in an inducing strategy, the product of Bob's observables
has eigenvalues $e^{i 2\mu}$ and $e^{-i 2\mu}$.

Therefore, the full correlation $P_{p,r}$ is induced by an ideal strategy
that can induce $P_{\hat{A}}$ and $P_{-\pi/p}$
for some odd prime $p$ whose primitive root is $r$. 
We can use $P_{-\pi/p}$ to certify that
the unitary $O$ has eigenvalues $\omega_p$ and $\omega_p^{-1}$.
Combining this observation with the fact that $O$ satisfies the relation \Cref{eq:uor},
we can prove that the correlation $P_{p,r}$ self-tests the state
$\ket{\tpsi}$ defined by
\begin{align*}
	\ket{\tpsi} = \frac{1}{\sqrt{p-1}} \sum_{j=1}^{p-1} \ket{j}\ket{p-j} .
\end{align*}
More precisely, we prove that if some quantum strategy using a shared state
$\ket{\psi}$ induces $P_{p,r}$, then there exist a local isometry
$\Phi_A \x \Phi_B$ and a state $\ket{junk}$ such that
\begin{align}
    \label{eq:distance}
    \Phi_A\x\Phi_B(\ket{\psi}) = \ket{junk}\x \ket{\tpsi}. 
\end{align}
Note that the isometry $\Phi_A \x \Phi_B$ only involves local operations
on Alice's and Bob's sides.
The isometry $\Phi_A \x \Phi_B$ captures the essence of the generalized 
swap-isometry proposed in \cite{yang2013}, but it is constructed in a different way.

The last step of proving Theorem \ref{thm:inf} involves a number theory result.
It has been shown that there exists $r \in \set{2, 3, 5}$
such that there are infinitely many primes whose smallest primitive is $r$ \cite{DRHeath,murty1988}.
The set $D$ in the statement of Theorem \ref{thm:inf} is the set of all such primes and
we can prove Theorem \ref{thm:inf} by applying the self-testing result of $P_{p,r}$ to all $p \in D$.

\subsection*{Structure of the paper}
We start with notations and background information in \cref{sec:prelim}.
We present a correlation that can enforce the relation $UOU\ct = O^r$
in \cref{sec:enforce} and present an extension of $P_{\mu}$ in \cref{sec:ext}. 
Then we define the correlation $P_{p,r}$ and prove its self-testing property in \cref{sec:self-test}.
Conclusions and open problems are discussed
in \cref{sec:conclude}. 

\section{Preliminaries}
\label{sec:prelim}
In this section, we introduce our notations and bipartite quantum correlations.
The EPR pair is denoted by 
\begin{align}
	\ket{EPR} = \frac{1}{\sqrt{2}}(\ket{00} + \ket{11}).
\end{align}
The $d$-th root of unity is denoted by $\omega_d:=e^{2\pi i/d}$. 
The set $\set{0,1,\ldots n-1}$ is denoted by $[n]$.

When we write 
operator relation with respect to a state, we use the following style.
Let $\calH_A$ and $\calH_B$ be two Hilbert spaces.
When it is clear from the context that 
unitaries $U_A$ and $U_B$ 
act on $\calH_A$ and $\calH_B$ respectively,
we write $U_A \x U_B \ket{\psi}$ as $U_AU_B \ket{\psi}$ ,
$U_A \x \1 \ket{\psi}$ as $U_A\ket{\psi}$, and $\1 \x U_B \ket{\psi}$
as $U_B\ket{\psi}$ for some $\ket{\psi} \in \calH_A \x \calH_B$.

To introduce \emph{bipartite quantum correlations}, we consider a \emph{nonlocal scenario} with two players, Alice and Bob. 
Each of them is requested
to give an answer for some question. 
Alice's question is chosen from the set $[n_A]$ and her answer should be from the set $[m_A]$.
Bob's question is chosen from the set $[n_B]$ and his answer should be from the set $[m_B]$.
Therefore, a nonlocal scenario is described by the tuple $([n_A], [n_B], [m_A], [m_B])$.
A \emph{bipartite correlation} of a nonlocal scenario
    $ ([n_A], [n_B], [m_A], [m_B])$
    is a function $P: [n_A] \times [n_B] \times [m_A] \times [m_B] \to \R_{\geq 0}:
    (i,j,k,l) \mapsto \pr{k,l}{i, j}$
    where $\pr{k, l}{i,j}$ is the probability for Alice
to answer $k$ and Bob to answer $l$ when the question to Alice 
is $i$ and to Bob is $j$.
The size of the correlation $P$ is $n_A n_B m_A m_B$.

Recall that
a set of self-adjoint operators on $\calH$, $\set{P_j \;|\;  j \in [n]}$,
is a \emph{projective measurement}
if $P_i^2 = P_i$ for all $i \in [n]$, 
$P_iP_j = 0$ for all $ i \neq j$, and 
$\sum_{j \in [n]} P_j = \1_{\calH}$.
Each $P_i$ is called a projector. A \emph{binary observable} 
on a Hilbert space $\calH$ is a self-adjoint operator $M$ such that
$M^2 = \1_{\calH}$.

A \emph{quantum projective measurement strategy} for a nonlocal scenario $([n_A], [n_B], [m_A], [m_B])$
is a tuple 
\begin{align*}
        (\ket{\psi} \in \calH_A \x \calH_B,
        &\set{ \set{ P_{i}^{(k)} | k \in [m_A]} | i \in [n_A]}, \\
        &\set{ \set{ Q_{j}^{(l)} |  l \in [m_B]} | j \in [n_B]}),
\end{align*}
where $\calH_A$ and $\calH_B$ are Hilbert spaces of arbitrary dimension, 
$\set{ \set{ P_{i}^{(k)} | k \in [m_A]} | i \in [n_A]}$
and $\set{ \set{ Q_{j}^{(l)} |  l \in [m_B]} | j \in [n_B]}$
are two sets of projective measurements on $\calH_A$ and $\calH_B$ respectively.
Note that the tensor product structure indicates that the
two parties cannot communicate with each other, 
which is the reason why we say such scenario is nonlocal.
This quantum strategy induces the bipartite quantum correlation
\begin{align}
	\pr{k,l}{i,j} = \bra{\psi} P_i^{(k)} \x Q_j^{(l)} \ket{\psi},
\end{align}
for each $(i,j,k,l) \in [n_A] \times [n_B] \times [m_A] \times [m_B]$.

Next we introduce a type of correlations associated with a binary linear system.
Let $Ax=0$ be an $m \times n$ binary linear system, where each row of $A$ has $\kappa$ nonzero entries and
each entry of $A$ is from $\Z_2$.
For each $i \in [m]$, we define
 \begin{align*}
     &I_i = \set{ j \in [n] \;| \;A(i,j) = 1 } \\
     &S_i = \set{ x \in \Z_2^{I_i} \cong \Z_2^\kappa \;|\; \sum_{j \in I_i} 
     x(j) \equiv  0 \pmod{2} },
\end{align*}
where the isomorphism between $\Z_2^{I_i}$ and $\Z_2^\kappa$ is implicit.

A correlation $P: [m] \times [n] \times \Z_2^\kappa \times \Z_2 \to \R$ is a \emph{perfect correlation}
of $Ax =0$ if 
\begin{align*}
	\sum_{x, y: x \in S_i,  x(j) =y} \pr{x,y}{i,j} = 1, 
\end{align*} 
for all $i \in [m]$ and $j \in I_i$.
The implication of a perfect correlation of $Ax = 0$ is summarized in the following lemma.
\begin{lemma}
	\label{lm:pft_cor}
	For an $m$-by-$n$ binary linear system $Ax = 0$ where each row of $A$
	has $\kappa$ nonzero entries,
	if a quantum strategy 
	\begin{align*}
		(&\ket{\psi} \in \calH_A \x \calH_B, \\
		&\set{ \set{ P_i^{(x)} | x \in \Z_2^\kappa} | i \in [m] }, \\
		&\set{ \set{ Q_j^{(y)} | y \in \Z_2} | j \in [n] })
	\end{align*}
    induces a perfect correlation of $Ax = 0$,
	then there exist a set of binary observables $\set{M_j | j \in [n]}$ 
	on $\calH_A$ and a set of binary observables $\set{N_j | j \in [n]}$
	on $\calH_B$ such that 
	\begin{align*}
		&M_j  N_j \ket{\psi} = \ket{\psi}
	\end{align*}
	for all $j \in [n]$,
	\begin{align*}
		\prod_{j \in I_i} M_j  \ket{\psi} = \prod_{j \in I_i}  N_j &\ket{\psi} = \ket{\psi} 
	\end{align*}
	for all $i \in [m]$, and 
	\begin{align*}
		(M_j M_k) \ket{\psi} = (M_k M_j) \ket{\psi}
	\end{align*}
	for all $i \in [m]$ and $j,k \in I_i$.
\end{lemma}
The proof of the lemma can be found in the proof of 
Theorem 4 of \cite{cleve2017} and in Section 3 of
\cite{slofstra2017} so we omit it here.

Another correlation that we are interested in is the correlation that gives 
the maximal violation of
the $\cot(\mu)$-weighted $\CHSH$ inequality defined in \Cref{eq:chsh}. 

We first give the inducing strategy of this correlation.
Let $\mu \in [-\pi, \pi)$.
Define
\begin{align*}
	&\tP_0^{(0)} = \ketbra{0}{0}, && \tP_0^{(1)} = \ketbra{1}{1}, \\
	&\tP_1^{(0)} = \frac{1}{2}(\ket{0}+\ket{1})(\bra{0}+\bra{1}), && \tP_1^{(1)} = \1 - \tP_1^{(0)},
\end{align*}
and
\begin{align*}
	\tQ_0^{(0)} = &(\cos(\frac{\mu}{2}) \ket{0} + \sin(\frac{\mu}{2})\ket{1})\\
	&\times (\cos(\frac{\mu}{2}) \bra{0} + \sin(\frac{\mu}{2})\bra{1}), \\
	\tQ_0^{(1)} = &\1 - \tQ_0^{(0)}, \\
	\tQ_1^{(0)} = &(\cos(\frac{\mu}{2}) \ket{0} - \sin(\frac{\mu}{2})\ket{1})\\
	&\times(\cos(\frac{\mu}{2}) \bra{0} - \sin(\frac{\mu}{2})\bra{1}), \\
	\tQ_1^{(1)} = &\1 - \tQ_1^{(0)}.
\end{align*}
\begin{definition}
	The correlation $P_\mu: [2] \times [2] \times [2] \times [2] \to \R$ is induced by the strategy
	\begin{align*}
		(\ket{EPR}, \set{\set{\tP_x^{(a)} | a \in [2]} | x \in [2]},\\
		 \set{\set{\tQ_y^{(b)} | b \in [2]} | y \in [2]}),
	\end{align*}
	such that $P_\mu( a, b | x, y ) = \bra{EPR}P_x^{(a)} Q_y^{(b)} \ket{EPR}$.
\end{definition}
The correlation $P_\mu$ is known as the optimal correlation of the $\cot(\mu)$-weighted $\CHSH$ inequality. 
The self-testing property of $P_\mu$ is summarized in the following Lemma,
which was first proved in \cite[Proposition A.$3$]{jed2017}.
\begin{lemma}
	For $\mu \in [-\pi, \pi)$,
	if a quantum strategy $(\ket{\psi}, \set{\set{P_x^{(a)} | a \in [2]} | x \in [2]},
		 \set{\set{Q_y^{(b)} | b \in [2]} | y \in [2]})$ induces $P_\mu$, then
	there exist a local isometry $\Phi = \Phi_A \x \Phi_B$ and an auxiliary state $\ket{junk}$  such that
	\begin{align*}
		 \Phi( P_x^{(a)} \x Q_y^{(b)} \ket{\psi}) = \ket{junk} \x (\tP_x^{(a)} \x \tQ_y^{(b)}) \ket{EPR} 
	\end{align*}
	for $x,y \in \set{-1, 0, 1}$ where the subscript $-1$ refers to the identity operator and $\tP_x^{(a)}, \tQ_y^{(b)}$ 
	for $x,y,a,b \in \set{0,1}$ are defined above. 
\end{lemma}
Instead of giving the full proof, we list some key relations of this proof that will be reused later.
From the projective measurements of the strategy, we get binary observables
$P_x := P_x^{(0)}-P_x^{(1)}$ and $Q_y := Q_y^{(0)}-Q_y^{(1)}$ for $x,y \in [2]$.
Define
\begin{align*}
	&Z_A := P_0, &
	&X_A := P_1,\\
	&Z_B := \frac{Q_0+Q_1}{2\cos{\mu}}, && X_B := \frac{Q_0-Q_1}{2\sin{\mu}},
\end{align*}
then
\begin{align}
	\label{eq:za-zb}& Z_A\ket{\psi} = Z_B\ket{\psi},\\
	\label{eq:xa-xb}&X_A\ket{\psi} = X_B \ket{\psi}, \\
	\label{eq:xazb}&X_A(\1+Z_B)\ket{\psi} = X_B(\1-Z_A) \ket{\psi},\\
	\label{eq:zaxb}&Z_A(\1+X_B)\ket{\psi} = Z_B(\1-X_A) \ket{\psi},\\
	\label{eq:zaxa}&Z_AX_A\ket{\psi} = -X_AZ_A \ket{\psi},\\
	\label{eq:zaxaxbzb}&X_AZ_A \ket{\psi} = -X_BZ_B \ket{\psi}.
\end{align}

\section{Enforcing $UOU^{-1} = O^r$}
\label{sec:enforce}
In this section, we show how to construct a binary linear system such that the perfect correlation
associated with it can enforce the relation $UOU^{-1} = O^r$. Since the construction relies heavily
on group presentations, we give the group theory background first.
For more contexts, please refer to \cite{rotman2012}.

\subsection{Group theory background}

Let $S$ be a set of letters. We denote by $\mathcal{F}(S)$ the \emph{free group generated by $S$}, 
which consists of all finite words made from $\set{s, s^{-1}| s \in S}$ such that no $ss^{-1}$ or $s^{-1}s$ appears as a substring for any $s$,
where $s^{-1}$ denotes the inverse of $s$. 
The group law is given by concatenation and cancellation.

\begin{definition}[Group presentation]
    Given a set $S$, let $\mathcal{F}(S)$ be the free group generated by $S$ and let $R$ be a subset of $\mathcal{F}(S)$.
    Then $\ip{S:R}$ denotes the quotient of $\mathcal{F}(S)$
    by the normal subgroup generated by $R$.
    If a group $G$ is isomorphic to $\ip{S:R}$,
    then we say $G$ 
    has a presentation $\ip{S:R}$. 
\end{definition}
If both sets $S$ and $R$ are finite and a group $G$ is defined by $\ip{S:R}$, 
then we say the group $G = \ip{S:R}$ is \emph{finitely-presented}.
The elements of $S$ are the \emph{generators} and the elements
of $R$ are the \emph{relations}.
A relation $r \in R$ is written as $r = e$
to convey its significance in the quotient group $G$.
There are two types of relations that we will work with.
Considering a subset of generators $\set{ s_j | 1 \leq j \leq n}  \subseteq S$,
the relations of the form $\Pi_{j=1}^n s_j  = e$, where $e$ is the identity element, are called 
\emph{linear relations};
and the relations of the form
$s_i s_j s_i^{-1} = s_k$ for some $i \neq j$ are called \emph{conjugacy relations}.   

There are two special types of groups that we will work with in this section. We give the definitions below.
\begin{definition}[Conjugacy group]
    \label{def:spe_conj_grp}
    Let $C \subseteq [n] \times [n] \times [n]$,
    and
    \begin{align*}
        G = \ip{\set{s_i | i \in [n]}  :
        &\set{s_i^2=e | i \in [n]} \cup \\
        &\set{ s_i s_j s_i = s_k | (i,j,k)\in C}}.
    \end{align*}
    We say a group is a conjugacy group if it has a presentation of this form. 
\end{definition}
A conjugacy group is a special case of what Slofstra defined as 
a linear-plus-conjugacy group (Definition $26$ of \cite{slofstra2017}), since a conjugacy group does not have linear relations. 
The other type of group is directly related to a linear system.
\begin{definition}[Solution group]
	\label{def:presentation}
	Let $Ax = 0$ be an $m \times n$ binary linear system. The solution group of this system
	is the group
	\begin{align*}
		\Gamma(A) := \ip{
		&\set{x_i | i \in [n]} :  \set{ x_j^2 = e | j \in [n] } \\
				&\cup \set{ \prod_{j \in I_i} x_j = e | i \in [m]}  \\
				&\cup \set{ x_l x_k = x_k x_l | k,l \in I_i \text{ for some } i\in[m]}
				}.
	\end{align*}
\end{definition}

A \emph{representation} of a group $G$ on a finite-dimensional Hilbert space $\calH$ is a group homomorphism from $G$ to the group of unitary operators on $\calH$, denoted by $\calU(\calH)$. 
The first reason that we study solution groups is that a representation of a solution group gives us a quantum strategy
that induces a perfect correlation of the linear system \cite[Theorem 18]{slofstra2017}.
The second reason is that, given a solution group, we can construct a binary linear system by converting a relation of the form $x_i x_j x_k = e$
into a linear equation $x_i + x_j + x_k = 0$. In the next subsection, we are going to construct a solution group first and then
extract the binary linear system from it.

Lastly, for groups $G$ and $K$, an \emph{embedding} of $G$ into $K$ is an injective group homomorphism $\phi: G \rightarrow K$.
The relation between conjugacy groups and solution groups is that any conjugacy group can be embedded in a solution group,
as proved in \cite[Proposition $27$]{slofstra2017}.

\subsection{Constructing the linear system}
This subsection is devoted to proving the following proposition.
\begin{proposition}
	\label{prop:hata}
	Let $r \geq 2$ be a positive integer.
	There exists a binary linear system $\hat{A}x = 0$ such that the 
	following holds.
	If a quantum strategy $S = (\ket{\psi} \in \calH_A\x\calH_B,
	\set{\set{P_x^{(a)}}}, \set{\set{Q_y^{(b)}}})$ induces
	a perfect correlation of $\hat{A}x =0$, then
	there exist binary observables $\set{M_{u_1}, M_{u_2},
	M_{o_1}, M_{o_2}}$ on $\calH_A$ and 
	$\set{N_{u_1}, N_{u_2}, N_{o_1}, N_{o_2}}$ on $\calH_B$
	such that
	\begin{align*}
		&M_{u_1}M_{u_2} (M_{o_1}M_{o_2}) M_{u_2}M_{u_1} \ket{\psi} 
		=   (M_{o_1}M_{o_2})^r \ket{\psi} \\
		&N_{u_1}N_{u_2} (N_{o_1}N_{o_2}) N_{u_2}N_{u_1} \ket{\psi} 
		=   (N_{o_1}N_{o_2})^r \ket{\psi}.
	\end{align*}
\end{proposition}
Note that in the statement of \Cref{prop:hata} $r$ can be any positive integer greater than 1, and it doesn't have to be from the set $\set{2,3,5}$.
To prove \Cref{prop:hata}, we need the following lemma to combine operator relations with
respect to the same shared state.
\begin{lemma}[Lemma $7$ of \cite{parallel_ghz}]
    \label{lm:sub}
    Let $\ket{\psi} \in \calH$ be a quantum state.
    Suppose there exist unitaries 
    $\set{V} \cup \set{V_i \;|\; i \in [k]} \cup \set{ M_i  \;|\; i \in [n]}$ on $\calH$ commuting with
    $\set{ N_i \;|\; i \in [n]}$ on $\calH$ such that
    \begin{align*}
        &M_i \ket{\psi} = N_i \ket{\psi}
    \end{align*}
    for each $i \in [n]$, and 
    \begin{align*}
        &V \ket{\psi} = \prod_{i \in [k]} V_i \ket{\psi}.
    \end{align*}
    Then,
    \begin{align*}
        V\prod_{i \in [n]} M_i \ket{\psi} 
        = \left(\prod_{i \in [k]} V_i\right) 
        \left(\prod_{i \in [n]} M_i \right)\ket{\psi}.
    \end{align*}
\end{lemma}
\begin{proof}[Proof of \cref{prop:hata}]
The linear system $\hat{A}x = 0$ is constructed from a solution group, wherein
the following group is embedded.
For $r \geq 2$, define 
\begin{align}
	G:=\ip{u, o: uou^{-1} = o^r}.
\end{align}
By \cite[Proposition $33$]{slofstra2017}, $G$ can be embedded into a conjugacy group
$G_c = \ip{S_c: R_c}$ where $S_c$ contains $\set{u_1, u_2, o_1, o_2}$
and $u_1^2 = u_2^2 = o_1^2 =o_2^2 = e$.
We also know that the embedding $\phi: G \to G_c$ maps $u$ to $u_1u_2$
and $o$ to $o_1o_2$. In other words, in $G_c$
\begin{align}
	\label{eq:uor_ext}
	u_1u_2 (o_1o_2) u_2 u_1 = (o_1o_2)^r.
\end{align}

By \cite[Proposition $27$]{slofstra2017}, $G_c$ can be embedded into a solution group $\Gamma(\hat{A}):= \ip{S_\Gamma, R_\Gamma}$.
Moreover, $\set{u_1, u_2, o_1, o_2} \subseteq S_\Gamma$ and the embedding 
$\phi': G_c \to \Gamma(\hat{A})$ maps $s$ to $s$ for each $s \in \set{u_1, u_2, o_1, o_2}$.
Therefore, $G$ is embedded in $\Gamma(\hat{A})$ and we get the
binary linear system $\hat{A}x =0$.

Since $G_c$ is embedded in $\Gamma(\hat{A})$, we know that 
the relation in \Cref{eq:uor_ext} can be reconstructed by
substituting in relations in $R_\Gamma$.
Then, the statement of the proposition follows from \cref{lm:pft_cor,lm:sub}.
\end{proof}
Following \cite[Remark 30]{slofstra2017}, we can get that $\hat{A}x = 0$ has
$n(r) := 16s+75$ variables and $m(r) := 14s + 62$ equations, where each equation has $3$ nonzero variables.
We assume that in this system $x_0$ corresponds to $o_1$ and $x_1$ corresponds to $o_2$.

Next we show that there exists a quantum strategy that can induce a perfect correlation of $\hat{A}x =0$.
The correlation is denoted by $P_{\hat{A}}$ and the strategy is denoted by $S_{\hat{A}}$, which is based on
a representation of $\Gamma(\hat{A})$.

We first give a representation of $G_c$.
Let $p$ be an odd prime number such that $r$ is a primitive root of $p$.
Note that here we don't require that $r$ is the smallest primitive root of $p$.
Let $\C^{p-1} = \spn( \set{ \ket{j } | 1 \leq j \leq p-1} )$.
A second basis of $\C^{p-1}$ is $\set{ \ket{x_j} | 1 \leq j \leq p-1}$,
where
\begin{align}
    \label{eq:ketj}
    &\ket{x_j} = -\frac{1}{\sqrt{2}}(\ket{j} + i\ket{p-j}), \\
    \label{eq:ketp-j}
    &\ket{x_{p-j}} = \frac{-\omega_{2p}^j}{\sqrt{2}}(\ket{j} - i\ket{p-j})	
\end{align}
for $1 \leq j \leq \frac{p-1}{2}$.
Note that another form of this basis is $\set{\ket{x_{r^j}} | j \in [p-1]}$, 
where the subscript $r^j$ is taken modulo $p$ implicitly.
Based on the second basis, we define a third basis of $\C^{p-1}$,
$\set{ \ket{u_k} | k \in [p-1]}$ defined by
\begin{align*}
	\ket{u_k} = \frac{1}{\sqrt{p-1}} \sum_{j=0}^{p-2} \omega_{p-1}^{jk} \ket{x_{r^j}}.
\end{align*}
On $\C^{p-1}$, we define
\begin{align}
	\label{eq:o1}
	O_1 = &\sum_{j=1}^{(p-1)/2} \omega_p^j \ketbra{x_j}{x_{p-j}} + \omega_p^{-j} \ketbra{x_{p-j}}{x_{j}}, \\
	\label{eq:o2}
	O_2 = &\sum_{j=1}^{(p-1)/2}  \ketbra{x_j}{x_{p-j}} + \ketbra{x_{p-j}}{x_{j}}, \\
	U_1 =& \ketbra{u_0}{u_0} -\ketbra{u_{(p-1)/2}}{u_{(p-1)/2}} \nonumber \\ 
    \label{eq:u1}+ &\sum_{k=1}^{(p-3)/2}\left( \omega_{p-1}^k\ketbra{u_k}{u_{p-1-k}} + \omega_{p-1}^{-k}\ketbra{u_{p-1-k}}{u_k}\right),\\
    U_2 =&\ketbra{u_0}{u_0} + \ketbra{u_{(p-1)/2}}{u_{(p-1)/2}} \nonumber \\
    \label{eq:u2} +&\sum_{k=1}^{(p-3)/2}\left(\ketbra{u_{p-1-k}}{u_k} + \ketbra{u_k}{u_{p-1-k}}\right).
\end{align}
It can be checked that
\begin{align*}
	&O_1O_2 = \sum_{j\in[p-1]} \omega_p^{r^j} \ketbra{x_{r^j}}{x_{r^j}},\\
	&U_1U_2 = \sum_{j \in [p-1]} \ketbra{x_{r^{j-1}}}{x_{r^j}}, \\
	&U_1U_2 (O_1O_2) U_2 U_1 = (O_1O_2)^r.
\end{align*}
Hence, we can follow the proof of \cite[Proposition $33$]{slofstra2017}
to extend $\rho: G_c \to \calU(\C^{p-1}) : u_1 \mapsto U_1, u_2 \mapsto U_2,
o_1 \mapsto O_1, o_2 \mapsto O_2$ to a representation of $G_c$,
still denoted by $\rho$.
Then, following the proofs of \cite[Proposition $27$ and Lemma $29$]{slofstra2017},
$\rho$ can be extended to a representation of $\Gamma(\hat{A})$, 
$\rho': \Gamma(\hat{A}) \to \calU(\C^{p-1} \x \C^2 \x \C^2)$.
In particular, for any $s \in \set{u_1, u_2, o_1, o_2}$, 
\begin{align*}
	\rho'(s) = \rho(s) \x \1_{\C^2} \x \1_{\C^2}.
\end{align*} 

Let 
\begin{align}
	\label{eq:tpsi}
	\ket{\tpsi} := \frac{1}{\sqrt{p-1}}\sum_{j=1}^{p-1} \ket{x_j} \ket{x_{p-j}},
\end{align}
and $\Pi_s^{(0)}, \Pi_s^{(1)}$ be the projectors onto the $+1$ and $-1$-eigenspaces
of $\rho'(s)$ for each $s \in S_\Gamma$.
Then we can construct a strategy
\begin{align*}
	S_{\hat{A}} = (&\ket{\tpsi}\x \ket{EPR}^{\x 2}, \\
	&\set{ \set{ P_i^{(x)} | x \in \Z_2^3} | i \in [\mr]},\\
	&\set{ \set{ \Pi_s^{(y)} | y \in \Z_2 } | s \in S_\Gamma}),
\end{align*}
where $P_i^{(x)}$ can be constructed from $\set{\Pi_s^{(0)}, \Pi_s^{(1)} | s \in S_\Gamma}$.
Note that since the variables of $\hat{A}$ are in one-to-one correspondence with the generators in $S_\Gamma$,
we label Bob's projectors by the corresponding generators.
\begin{definition}
The correlation $P_{\hat{A}} : [\mr] \times S_\Gamma \times \Z_2^3 \times \Z_2 \to R$ is defined
by
\begin{align*}
	&P_{\hat{A}}( x, y | i, s ) \\
	=&[\bra{\tpsi} \x \bra{EPR}^{\x 2}] (P_i^{(x)} \Pi_s^{(y)}) [\ket{\tpsi}\x \ket{EPR}^{\x 2}].
\end{align*}
\end{definition}
It can be checked that $P_{\hat{A}}$ is a perfect correlation of $\hat{A}x = 0$.

\section{Extending the correlation $P_\mu$}
\label{sec:ext}

\subsection{Revisiting the self-testing property of $P_\mu$}
In this section, we show how to use $P_\mu$ to argue the eigenvalue
of an unknown unitary.
\begin{proposition}
	\label{prop:pmu}
	For $\mu \in [-\pi, \pi)$,
	if a quantum strategy $(\ket{\psi} \in \calH_A\x\calH_B, \set{\set{P_x^{(a)} | a \in [2]} | x \in [2]},
		\set{\set{Q_y^{(b)} | b \in [2]} | y \in [2]})$ induces $P_\mu$, then
	there exist quantum states $\ket{\psi_1}, \ket{\psi_2} \in \calH_A\x\calH_B$
	such that 
	\begin{align*}
		&Q_0Q_1\ket{\psi_1} = e^{-i2\mu} \ket{\psi_1},\\
		&Q_0Q_1\ket{\psi_2} = e^{i2\mu} \ket{\psi_2},
	\end{align*}
	where $Q_y := Q_y^{(0)} - Q_y^{(1)}$
	for $y \in[2]$.
\end{proposition}
\begin{proof}
    Let $P_x = P_x^{(0)} - P_x^{(1)}$
    for $x \in \set{0,1}$.
	The states in \Cref{prop:pmu} are 
	\begin{align}
    		\label{eq:psi1}
		\ket{\psi_1} =
		(P_{0}^{(0)} +  iP_1 P_{0}^{(1)}) \ket{\psi}, \\
		\label{eq:psi2}
		\ket{\psi_2} = 
		(P_{0}^{(0)} -  iP_1 P_{0}^{(1)}) \ket{\psi}.
	\end{align}
	We first show that $\norm{\ket{\psi_1}} = 1$.
	\begin{align*}
		&\norm{\ket{\psi_1}}^2 \\
		=&\bra{\psi}(P_{0}^{(0)}+P_{0}^{(1)}-iP_{0}^{(1)}P_{1}P_{0}^{(0)}+iP_{0}^{(0)}P_{1}P_{0}^{(1)}\ket{\psi}\\
		=&1 - i \bra{\psi}P_{0}^{(1)}P_{1}P_{0}^{(0)}-P_{0}^{(0)}P_{1}P_{0}^{(1)}\ket{\psi}. 
	\end{align*}
	Recall that $Z_A = P_0$, $X_A = P_1$, $Z_B = (Q_0+Q_1)/2\cos(\mu)$ and
	$X_B = (Q_0 - Q_1)/2\sin(\mu)$. Using \cref{eq:xa-xb,eq:zaxaxbzb}, we know
	\begin{align*}
		P_{1}P_{0}^{(0)} \ket{\psi} =& \frac{X_A (\1+Z_A)}{2}\ket{\psi} \\
		=& \frac{X_B(\1 - Z_B)}{2} \ket{\psi} \\
		=& \frac{(\1 + Z_B)X_B}{2} \ket{\psi} \\
		P_0^{(1)} \ket{\psi} = & \frac{\1 - Z_A}{2} \ket{\psi} \\
		=& \frac{\1 -Z_B}{2} \ket{\psi},
	\end{align*}
	so $\bra{\psi} P_0^{(1)}P_{1}P_{0}^{(0)} \ket{\psi} = 0$.
	With similar reasoning, we get $\bra{\psi} P_0^{(0)}P_{1}P_{0}^{(1)} \ket{\psi} = 0$.
	Therefore, $\norm{\ket{\psi_1}} = 1$. The derivation of $\norm{\ket{\psi_2}} = 1$ is very similar,
	so we omit it here.
	
	Next, we show $Q_0Q_1\ket{\psi_1} = e^{-i2\mu} \ket{\psi_1}$ and
	$Q_0Q_1\ket{\psi_2} = e^{i2\mu} \ket{\psi_2}$.
	From \cref{eq:za-zb}, we get that 
	\begin{align*}
		Z_BP_0^{(0)} \ket{\psi} =& \frac{Z_B (\1 + Z_A)}{2} \ket{\psi} \\
		=&\frac{Z_B + \1}{2} \ket{\psi} \\
		=&\frac{\1 + Z_A}{2} \ket{\psi} \\
		=&P_0^{(0)} \ket{\psi}.
	\end{align*}
	With similar reasoning, we get
	\begin{align*}
		Z_B P_0^{(1)} \ket{\psi} = - P_0^{(1)}\ket{\psi}.
	\end{align*}
	Substituting the expression of $Z_B$, we see that 
	\begin{align*}
		&(Q_0 + Q_1) P_0^{(0)} \ket{\psi} = 2\cos(\mu) P_0^{(0)} \ket{\psi}, \\
		&(Q_0 + Q_1) P_0^{(1)} \ket{\psi} = - 2\cos(\mu) P_0^{(1)} \ket{\psi}.
	\end{align*}
	From \cref{eq:xa-xb,eq:zaxa,eq:za-zb,eq:xazb}, we get that 
	\begin{align*}
		&X_B P_0^{(0)} \ket{\psi} = X_A P_0^{(1)} \ket{\psi}, \\
		&X_B P_0^{(1)} \ket{\psi} = X_A P_0^{(0)} \ket{\psi}.	
	\end{align*}
	Substituting in the expression of $X_B$, we get that 
	\begin{align*}
		&(Q_0 -Q_1) P_0^{(0)} \ket{\psi} = 2\sin(\mu) P_1 P_0^{(1)} \ket{\psi}, \\
		&(Q_0 -Q_1) P_0^{(1)} \ket{\psi} = 2\sin(\mu) P_1 P_0^{(0)} \ket{\psi}.
	\end{align*}
	Simple calculation gives us that
	\begin{align*}
		&Q_0 P_0^{(0)} \ket{\psi} = [\cos(\mu) P_0^{(0)} + \sin(\mu) P_1 P_0^{(1)}] \ket{\psi}, \\
		&Q_1 P_0^{(0)} \ket{\psi} = [\cos(\mu) P_0^{(0)} - \sin(\mu) P_1 P_0^{(1)}] \ket{\psi}, \\
		&Q_0 P_0^{(1)} \ket{\psi} = [-\cos(\mu) P_0^{(1)} + \sin(\mu) P_1 P_0^{(0)}] \ket{\psi}, \\
		&Q_1 P_0^{(1)} \ket{\psi} = [-\cos(\mu) P_0^{(1)} - \sin(\mu) P_1 P_0^{(0)}] \ket{\psi}.
	\end{align*}
	Then,
	\begin{align*}
		&Q_0 Q_1 P_0^{(0)} \ket{\psi} =  
		 [\cos(2\mu) P_0^{(0)} + \sin(2\mu)P_1 P_0^{(1)}] \ket{\psi}, \\
		 &Q_0Q_1 P_1 P_0^{(1)} \ket{\psi} =
		 [\cos(2\mu) P_1P_0^{(1)} -\sin(2\mu)P_0^{(0)}]\ket{\psi}.
	\end{align*}
	Finally, the fact that $\ket{\psi_1}$ and $\ket{\psi_2}$ are $e^{-2i\mu}$ and
	$e^{2i\mu}$ eigenvectors pf $Q_0Q_1$
	follows \Cref{eq:psi1,eq:psi2}. 
\end{proof}
\subsection{The correlation $\hat{P}_{-\pi/p}$}
In the rest of the work, we fix $\mu = - \pi/p$ for some odd prime $p$.
We will introduce a correlation that is extended from $P_{-\pi/p}$ and the correlation is
denoted by $\hat{P}_{-\pi/p}$. We define $\hat{P}_{-\pi/p}: [5] \times [5] \times [3] \times [3] \to \R$ by defining its
inducing quantum strategy.

In $\C^{p-1}$, we define a subspace $V = \spn(\set{\ket{1}, \ket{p-1}})$ and we denote 
the projector onto $V$ by $\Pi_V$.
For question $x = 0$, let
\begin{align*}
	\overline{P}_0^{(a)} = \overline{Q}_0^{(a)} = \begin{cases}
	\Pi_V & \text{ if } a = 0\\
	\1 - \Pi_V & \text{ if } a = 1 \\
	0 & \text{ otherwise. }
	\end{cases}
\end{align*}
For questions $x = 1, 2$, let $\Lambda_{O_x}^{(0)}, \Lambda_{O_x}^{(1)}$
be the projectors onto the $+1$- and $-1$-eigenspaces of $O_x$,
where $O_1$ and $O_2$ are defined in \Cref{eq:o1,eq:o2},
and let
\begin{align*}
	\overline{P}_x^{(a)} = \overline{Q}_x^{(a)} = \begin{cases}
	\Lambda_{O_x}^{(0)} & \text{ if } a = 0\\
	\Lambda_{O_x}^{(1)} & \text{ if } a = 1 \\
	0 & \text{ otherwise. }
	\end{cases}
\end{align*}
For question $x = 3$, let
\begin{align*}
	\overline{P}_3^{(a)} = \overline{Q}_3^{(a)} = \begin{cases}
	\ketbra{1}{1} & \text{ if } a = 0\\
	\ketbra{p-1}{p-1} & \text{ if } a = 1 \\
	\1- \Pi_V & \text{ otherwise. }
	\end{cases}
\end{align*}
For question $x = 4$, let
\begin{align*}
	\overline{P}_4^{(a)} = \overline{Q}_4^{(a)} = \begin{cases}
	\frac{(\ket{1}+\ket{p-1})(\bra{1}+\bra{p-1}}{2}) & \text{ if } a = 0\\
	\frac{(\ket{1}-\ket{p-1})(\bra{1}-\bra{p-1}}{2}) & \text{ if } a = 1 \\
	\1- \Pi_V & \text{ otherwise. }
	\end{cases}
\end{align*}
Substituting \Cref{eq:ketj,eq:ketp-j} into \Cref{eq:tpsi}, we get
\begin{align*}
	\ket{\tpsi} = \frac{1}{\sqrt{p-1}} \sum_{j=1}^{(p-1)/2} \omega_{2p}^j(\ket{j}\ket{j} + \ket{p-j}\ket{p-j}).
\end{align*} 
The inducing strategy is
\begin{align*}
	S_{-\pi/p} = (\ket{\tpsi}, \set{\set{ \overline{P}_x^{(a)} | a \in [3]} | x \in [5]}, \\
	\set{\set{ \overline{Q}_y^{(b)} | b \in [3]} | y \in [5]}).
\end{align*}
\begin{definition}
	The correlation $\hat{P}_{-\pi/p} : [5] \times [5] \times [3] \times [3] \to \R$
	is induced by $S_{-\pi/p}$:
	\begin{align*}
		\hat{P}_{-\pi/p}( a, b | x, y) = \bra{\tpsi} \overline{P}_x^{(a)} \overline{Q}_y^{(b)} \ket{\tpsi}.
	\end{align*}
\end{definition}
As an analogue of \cref{prop:pmu}, the implication of $\hat{P}_{-\pi/p}$ is summarized
in the next proposition.
\begin{proposition}
	\label{prop:pmu_ext}
	Let $(\ket{\psi} \in \calH_A\x\calH_B, \set{\set{P_x^{(a)} | a \in [3]} | x \in [5]},
		 \set{\set{Q_y^{(b)} | b \in [3]} | y \in [5]})$ 
		 be a strategy that
		 induces $\hat{P}_{-\pi/p}$, and
	let 
	\begin{align}
	    \label{eq:psi1}
	    \ket{\psi_1} =&\frac{1}{2}(P_{3}^{(0)} + iP_{4}P_{3}^{(1)} - iP_{4}P_{3}^{(0)} +P_{3}^{(1)}) \ket{\psi},
	\end{align}
	where $P_4 = P_4^{(0)} - P_4^{(1)}$.
	Then,
	$\norm{\ket{\psi_1}}^2 = 1/(p-1)$ and 
	\begin{align}
		\label{eq:p1p2} &P_1P_2 \ket{\psi_1} = \omega_p^{-1} \ket{\psi_1} \\
		\label{eq:q1q2} &Q_1Q_2\ket{\psi_1} = \omega_p \ket{\psi_1},
	\end{align}
	where $P_x := P_x^{(0)} - P_x^{(1)}$ and $Q_y := Q_y^{(0)} - Q_y^{(1)}$
	for $x,y \in \set{1, 2}$.
\end{proposition}
To help with the proof of this proposition, we first give some values of $\hat{P}_{-\pi/p}$.
\begin{align*}
	&\hat{P}_{-\pi/p}(a,b| 0, 0) = 
	\begin{cases}
		2/(p-1) &\text{ if } a = b = 0 \\
		(p-3)/(p-1) & \text{ if } a = b = 1 \\
		0 & \text{ otherwise. }
	\end{cases} \\
	&\hat{P}_{-\pi/p}(a,b| 3, 3) = 
	\begin{cases}
		1/(p-1) &\text{ if } a = b = 0 \\
		1/(p-1) & \text{ if } a = b = 1 \\
		(p-3)/(p-1) & \text{ if } a = b = 2 \\
		0 & \text{ otherwise. }
	\end{cases} \\
	&\hat{P}_{-\pi/p}(a,b| 0, 3) = 
	\begin{cases}
		1/(p-1) &\text{ if } a = 0, b \in [2]  \\
		(p-3)/(p-1) & \text{ if } a =1 , b = 2 \\
		0 & \text{ otherwise. }
	\end{cases}
\end{align*}

\begin{table}[H]
\begin{center}
\tabcolsep=0.06cm
\begin{tabular}{|c|c||c|c|c|c|}
\hline
\multicolumn{2}{|c|}{} &
\multicolumn{2}{|c|}{$x=3$}&
\multicolumn{2}{|c|}{$x=4$} \\
\cline{3-6}
\multicolumn{2}{|c|}{} &
$a = 0$ & $a=1$ &
$a = 0$ & $a=1$ \\
\hline
\hline
\multirow{2}{*}{$y = 1$} & $b=0$ & $\frac{\cos^2(\pi/2p)}{p-1}$ & $\frac{\sin^2(\pi/2p)}{p-1}$  
& $\frac{1-\sin(\pi/p)}{2(p-1)}$ & $\frac{1+\sin(\pi/p)}{2(p-1)}$  \\
\cline{2-6}
&$b=1$ & $\frac{\sin^2(\pi/2p)}{p-1}$ & $\frac{\cos^2(\pi/2p)}{p-1}$ 
&  $\frac{1+\sin(\pi/p)}{2(p-1)}$ & $\frac{1-\sin(\pi/p)}{2(p-1)}$ \\
\hline
\multirow{2}{*}{$y = 2$} & $b=0$ & $\frac{\cos^2(\pi/2p)}{p-1}$ & $\frac{\sin^2(\pi/2p)}{p-1}$  & 
$ \frac{1+\sin(\pi/p)}{2(p-1)}$ & $ \frac{1-\sin(\pi/p)}{2(p-1)}$   \\
\cline{2-6}
&$b=1$ & $\frac{\sin^2(\pi/2p)}{p-1}$ & $\frac{\cos^2(\pi/2p)}{p-1}$ &   
$ \frac{1-\sin(\pi/p)}{2(p-1)}$ & $ \frac{1+\sin(\pi/p)}{2(p-1)}$  \\
\hline
\end{tabular}
\end{center}
\caption{The correlation for $x \in \set{3, 4}$, $y \in \set{1,2}$ and $a,b \in [2]$.}
\label{tb:chsh}
\end{table}
\begin{proof}
	From the definition of $\hat{P}_{-\pi/p}$, it is easy to see that
	\begin{align*}
		P_x^{(2)} \ket{\psi} = Q_x^{(2)} \ket{\psi} = 0
	\end{align*}
	for $x,y \in [3]$.
	Then
	\begin{align*}
		P_x^2 \ket{\psi} = &[P_x^{(0)} + P_x^{(1)}] \ket{\psi} + P_x^{(2)} \ket{\psi}\\
		 =& \ket{\psi}
	\end{align*}
	for $x \in \set{1, 2}$. Similarly, we see that $Q_y^2 \ket{\psi} = \ket{\psi}$ for $y \in \set{1,2}$.
	
	Next, we will show
	\begin{align*}
	  P_0^{(0)} \ket{\psi} &= 
	  (P_{3}^{(0)}+P_{3}^{(1)})\ket{\psi}\\
	  &= (P_{4}^{(0)}+P_{4}^{(1)})\ket{\psi} \\
	  &=Q_0^{(0)} \ket{\psi}\\
	  &= (Q_{3}^{(0)}+Q_{3}^{(1)})\ket{\psi}\\
	  &= (Q_{4}^{(0)}+Q_{4}^{(1)})\ket{\psi}\\
 	P_0^{(1)} \ket{\psi} 
	&= P_{3}^{(2)} \ket{\psi}
	= P_{4}^{(2)} \ket{\psi} =\\
	Q_0^{(1)} \ket{\psi}
	&= Q_{3}^{(2)} \ket{\psi}
	= Q_{4}^{(2)} \ket{\psi},
\end{align*}
and
\begin{align*}
&P_{3}^{(0)} \ket{\psi} = Q_{3}^{(0)} \ket{\psi} && P_{3}^{(1)}\ket{\psi} = Q_{3}^{(1)} \ket{\psi}\\
&P_{4}^{(0)} \ket{\psi} = Q_{4}^{(0)} \ket{\psi} && P_{4}^{(1)}\ket{\psi} = Q_{4}^{(1)} \ket{\psi}.
\end{align*}
Instead of giving the full proof, we show $P_0^{(0)} \ket{\psi} = Q_0^{(0)} \ket{\psi}$ to demonstrate
	  the ideas.
	  From the correlation, we know $\bra{\psi} P_0^{(0)}Q_0^{(0)} \ket{\psi} = 
	  \bra{\psi} P_0^{(0)} \ket{\psi} = \bra{\psi} Q_0^{(0)} \ket{\psi} = 2/(p-1)$.
	  Then,
	  \begin{align*}
	  	&\norm{P_0^{(0)}\ket{\psi} - Q_0^{(0)}\ket{\psi}}^2 \\
		=& \bra{\psi} P_0^{(0)} \ket{\psi} + \bra{\psi} Q_0^{(0)} \ket{\psi}
		- 2\bra{\psi} P_0^{(0)}Q_0^{(0)} \ket{\psi} \\
		=&0,
	  \end{align*}
	  which implies that $P_0^{(0)} \ket{\psi} = Q_0^{(0)}\ket{\psi}$.

Then, we can show that $\hat{P}_{-\pi/p}$ can be "reduced" to $P_{-\pi/p}$
by proving that
\begin{align*}
	S = \Big(\frac{P_0^{(0)}\ket{\psi}}{\norm{P_0^{(0)}\ket{\psi}}}, 
	&\set{\set{P_x^{(0)}, P_x^{(1)}} | x \in \set{3,4}}, \\
	&\set{\set{Q_y^{(0)}, Q_y^{(1)} } | y \in \set{1, 2}}\Big)
\end{align*} 
can induce $P_{-\pi/p}$, and that
\begin{align*}
	S' = \Big(\frac{P_0^{(0)}\ket{\psi}}{\norm{P_0^{(0)}\ket{\psi}}}, 
	&\set{\set{P_x^{(0)}, P_x^{(1)}} | x \in \set{1,2}}, \\
	&\set{\set{Q_y^{(0)}, Q_y^{(1)} } | y \in \set{3, 4}}\Big)
\end{align*} 
can induce $P_{-\pi/p}$ with Alice and Bob's roles flipped. To prove $S$ can induce $P_{-\pi/p}$, we need to examine the terms of the form
$\bra{\psi} P_0^{(0)} P_{x}^{(a)} Q_y^{(b)} P_0^{(0)}\ket{\psi}$ for $x =3,4$, $y=1,2$ and $a,b = 0,1$.
We find that these terms relate to $\bra{\psi} P_x^{(a)} Q_y^{(b)} \ket{\psi}$ by
\begin{align*}
	   &\bra{\psi} P_x^{(a)} Q_y^{(b)} \ket{\psi} \\
	= &\bra{\psi}(P_0^{(0)} + P_0^{(1)}) P_x^{(a)} Q_y^{(b)} (P_0^{(0)} + P_0^{(1)})\ket{\psi} \\
	=&\bra{\psi}P_0^{(0)}P_x^{(a)} Q_y^{(b)}  P_0^{(0)}\ket{\psi},
\end{align*}
where we use the facts that $P_x^{(a)} P_0^{(1)} \ket{\psi} = P_x^{(a)} P_x^{(2)} \ket{\psi} = 0$
for the relevant values of $(x,y,a,b)$.
Therefore,
\begin{align*}
	\frac{\bra{\psi}P_0^{(0)}P_x^{(a)} Q_y^{(b)}  P_0^{(0)}\ket{\psi}}{\norm{P_0^{(0)}\ket{\psi}}^2} 
	= \frac{ \bra{\psi} P_x^{(a)} Q_y^{(b)} \ket{\psi} }{ \norm{P_0^{(0)}\ket{\psi}}^2},
\end{align*}
for the relevant values of $(x,y,a,b)$, and it is easy to verify that $S$ induces $P_{-\pi/p}$.
The proof of $S'$ induces $P_{-\pi/p}$ with Alice and Bob's roles flipped is similar, so we omit it here.

The state $\ket{\psi_1}$ can also be written as 
\begin{equation}
    \label{eq:psi1_form2}
	\ket{\psi_1} =\frac{1}{2} (\1 - iP_4)(P_{3}^{(0)} + iP_4 P_3^{(1)}) \ket{\psi}.
\end{equation}
The derivation of $\norm{\ket{\psi_1}}$ is very similar to the corresponding part in
the proof of \cref{prop:pmu}, so we omit it here.
Since $S$ can induce $P_{-\pi/p}$, by $\cref{prop:pmu}$, we know that 
\begin{align*}
	&Q_1Q_2 (P_{3}^{(0)} + iP_4 P_3^{(1)}) P_0^{(0)} \ket{\psi}\\
	 = &\omega_p (P_{3}^{(0)} + iP_4 P_3^{(1)}) P_0^{(0)} \ket{\psi}.
\end{align*}
On the other hand,
\begin{align*}
	&(P_{3}^{(0)} + iP_4 P_3^{(1)}) P_0^{(0)} \ket{\psi}\\
	 = &(P_{3}^{(0)} + iP_4 P_3^{(1)}) (P_3^{(0)}+P_3^{(1)})\ket{\psi}\\
	 = &(P_{3}^{(0)} + iP_4 P_3^{(1)}) \ket{\psi}.
\end{align*} 
Hence, using the fact that $Q_1Q_2$ commutes with $(\1 - iP_4)$, we know
\begin{align*}
	Q_1Q_2 \ket{\psi_1} = \omega_p \ket{\psi_1}.
\end{align*}

What remains to be proved is $P_1P_2 \ket{\psi_1} = \omega_p^{-1} \ket{\psi_1}$.
In order to prove it, we need another form of $\ket{\psi_1}$, which is
\begin{equation}
    \label{eq:psi1_q}
    \begin{aligned}
	\ket{\psi_1} =&\frac{1}{2} (Q_{3}^{(0)} - iQ_{4}Q_{3}^{(1)}+iQ_{4}Q_{3}^{(0)} + Q_{3}^{(1)})\ket{\psi}.\\
	=&(\1 +iQ_4)(Q_{3}^{(0)} - iQ_{4}Q_{3}^{(1)})\ket{\psi}, 
\end{aligned}
\end{equation}
where $Q_4 := Q_4^{(0)} - Q_4^{(1)}$.
Comparing the two forms of $\ket{\psi_1}$, it suffices to show
\begin{align*}
	P_{4}(P_{3}^{(1)} - P_{3}^{(0)}) \ket{\psi} = Q_4(Q_{3}^{(0)} - Q_{3}^{(1)}) \ket{\psi}.
\end{align*}
This equation can be derived in the following way
\begin{align*}
	&P_{4}(P_{3}^{(1)} - P_{3}^{(0)}) \ket{\psi} \\
	=&P_{4}(P_{3}^{(1)} - P_{3}^{(0)}) P_0^{(0)} \ket{\psi}  \\
	=&(Q_{3}^{(1)} - Q_{3}^{(0)})Q_4 Q_0^{(0)} \ket{\psi} \\
	=&Q_4 (Q_{3}^{(0)} - Q_{3}^{(1)})Q_0^{(0)} \ket{\psi} \\
	=&Q_4 (Q_{3}^{(0)} - Q_{3}^{(1)}) \ket{\psi},
\end{align*}
where we use the fact that \Cref{eq:zaxa} is satisfied in the inducing strategies $S$ and $S'$.
In the end, we apply \cref{prop:pmu} to $S'$ to see that
\begin{align*}
	&P_1P_2 \ket{\psi_1} \\
	 = &(\1 +iQ_4) P_1P_2(Q_3^{(0)} - iQ_4Q_3^{(1)})\ket{\psi}\\
	 =&\omega_p^{-1} \ket{\psi_1},
\end{align*}
which completes the proof.
\end{proof}

\section{The self-testing correlation}
\label{sec:self-test}
In this section, we introduce the correlation $P_{p,r}: [\nr+\mr+3] \times [\nr+\mr+3] \times [8] \times [8] \to \R$,
which can be thought of as the combination of $P_{\hat{A}}$ and $\hat{P}_{-\pi/p}$.

We define $P_{p,r}$ by giving the inducing strategy, which is based on $S_{\hat{A}}$ of
$P_{\hat{A}}$ and $S_{-\pi/p}$ of $\hat{P}_{-\pi/p}$.
Let
\begin{widetext}
\begin{align*}
	&\tP_x^{(a)} = 
	\begin{cases}
		\Pi_x^{(a)} &\text{ if } x \in [\nr] \text{ and } a \in [2], \\
		P_{x-\nr}^{(a)} & \text{ if } \nr \leq x \leq \nr+\mr-1,  \\
		\overline{P}_0^{(a)} \x \1_{\C^2}^{\x 2}& \text{ if } x = \nr+\mr \text{ and } a \in [2], \\
		\overline{P}_{x+2 -\nr-\mr}^{(a)} \x \1_{\C^2}^{\x 2} &\text{ if } x = \nr+\mr+1, \nr+\mr+2 \text{ and } a \in [3], \\
		0 & \text{ otherwise;}
	\end{cases} \\
	&\tQ_y^{(b)} = 
	\begin{cases}
		\Pi_y^{(b)} &\text{ if } y \in [\nr] \text{ and } b \in [2],\\
		P_{y-\nr}^{(b)} & \text{ if } \nr \leq y \leq \nr+\mr-1, \\
		\overline{Q}_0^{(b)} \x \1_{\C^2}^{\x 2}& \text{ if } y = \nr+\mr \text{ and } b \in [2], \\
		\overline{Q}_{y+2 -\nr-\mr}^{(b)} \x \1_{\C^2}^{\x 2} &\text{ if } y = \nr+\mr+1, \nr+\mr+2 \text{ and } b \in [3], \\
		0 & \text{ otherwise. }
	\end{cases}
\end{align*}
The inducing strategy is 
\begin{align*}
	\tilde{S} = (\ket{\tpsi} \x \ket{EPR}^{\x 2},
	 \set{\set{\tP_x^{(a)} | a \in [8]} | x \in [\nr+\mr+3]}, 
	 \set{ \set{ \tQ_y^{(b)} | b \in [8]} | y \in [\nr+\mr+3]}).
\end{align*}
\end{widetext}

Note that in the definition of $\tilde{S}$, a bijection between $[\nr]$ and $S_\Gamma$ is implicit.
\begin{definition}
	The correlation $P_{p,r}: [\nr+\mr+3] \times [\nr+\mr+3] \times [8] \times [8] \to \R$
	is induced by $\tilde{S}$ as
	\begin{align*}
		&P_{p,r}(a,b|x,y) \\
		=& [\bra{\tpsi} \x \bra{EPR}^{\x 2}] \tP_x^{(a)} \tQ_y^{(b)} [\ket{\tpsi} \x \ket{EPR}^{\x 2}].
	\end{align*}
\end{definition}
It is not hard to see that 
\begin{itemize}
\item When $x \in [\nr]$, $\nr \leq y \leq \nr+\mr-1$ and $a \in [2]$,
\begin{align*}
	P_{p,r}(a,b|x,y) = P_{\hat{A}}( b, a | y -\nr, x).
\end{align*}
\item When $\nr \leq x \leq  \nr+\mr-1$, $y \in [\nr]$ and $b \in [2]$,
\begin{align*}
	P_{p,r}(a,b|x,y) = P_{\hat{A}}( a, b | x-\nr, y).
\end{align*}
\item When $ x = y \in [\nr]$,
\begin{align*}
	P_{p,r}(0,0|x,x) + P_{p,r}(1,1|x,x) =1.
\end{align*}
\item When $x, y \in \set{0, 1, \nr+\mr, \nr+\mr+1, \nr+\mr+2}$,
define $f: \set{0, 1, \nr+\mr, \nr+\mr+1, \nr+\mr+2} \to [5]$ by
\begin{align*}
	f(x) = \begin{cases}
	x+1  \text{ if } x =0,1 , \\
	x - \nr-\mr  \text{ if } x = \nr+\mr, \\
	x+2 - \nr-\mr  \text{ otherwise.}
	\end{cases}
\end{align*}
Then, if $a, b \in [3]$
\begin{align*}
	P_{p,r}(a,b|x,y) = \hat{P}_{-\pi/p} ( a, b| f(x), f(y) ).
\end{align*}
\end{itemize}
\begin{proposition}
\label{prop:decomp_psi}
If a strategy 
\begin{align*}
	S = (&\ket{\psi} \in \calH_A \x \calH_B, \\
	&\set{\set{P_x^{(a)} | a \in [8]}| x \in [\nr+\mr+3]},\\
	&\set{\set{Q_y^{(b)} | b \in [8]}| y \in [\nr+\mr+3]})
\end{align*}
 induces $P_{p,r}$, then there exist unitaries $U_A \in \calU(\calH_A)$ 
and $U_B \in \calU(\calH_B)$ and a subnormalized state $\ket{\psi_1} \in \calH_A\x\calH_B$ such that 
\begin{align*} 
	& \norm{\ket{\psi_1}}^2 = \frac{1}{p-1} \\
	&\ket{\psi} = \sum_{j=1}^{(p-1)} (U_A\ct U_B\ct)^{\log_r j} \ket{\psi_1}
\end{align*}
where $\log_r j$ is the discrete log.
\end{proposition}
Before we prove the proposition, we fix some notations used
in the proofs of this proposition and the next theorem.
We relabel $P_{0}^{(a)}$ and $Q_{0}^{(b)}$ as $P_{o_1}^{(a)}$ and $Q_{o_1}^{(b)}$,
and relabel $P_{1}^{(a)}$ and $Q_{1}^{(b)}$ as $P_{o_2}^{(a)}$ and $Q_{o_2}^{(b)}$.
We can also identify variables of $\hat{A}x=0$ corresponding to the generators $u_1$ and $u_2$,
so we label the corresponding Alice and Bob's projectors as $P_{u_1}^{(a)}, P_{u_2}^{(a)}$
and $Q_{u_1}^{(b)}, Q_{u_2}^{(b)}$.
For $s \in \set{u_1, u_2, o_1, o_2}$, we further define
\begin{align*}
	P_{s} := P_{s}^{(0)} - P_{s}^{(1)} && Q_s=Q_{s}^{(0)} - Q_{s}^{(1)},
\end{align*}	
and
\begin{align*}
	&O_A := P_{o_1}P_{o_2} && O_B := Q_{o_1}Q_{o_2}, \\
	&U_A :=  P_{u_1}P_{u_2} && U_B := Q_{u_1}Q_{u_2}.
\end{align*}
\begin{proof}
Define $P_{\nr+\mr+1} := P_{\nr+\mr+1}^{(0)} - P_{\nr+\mr+1}^{(1)}$,
$P_{\nr+\mr+2} := P_{\nr+\mr+2}^{(0)} - P_{\nr+\mr+2}^{(1)}$, and
\begin{align*}
\ket{\psi_1} = &
\frac{1}{2}(P_{\nr+\mr+1}^{(0)} +P_{\nr+\mr+1}^{(1)} \\
 &-iP_{\nr+\mr+2}P_{\nr+\mr+1}) \ket{\psi}
\end{align*}
Since $S$ can induce $\hat{P}_{-\pi/p}$, by \cref{prop:pmu_ext}, we know
$\norm{\ket{\psi_1}} = 1/\sqrt{p-1}$, and 
\begin{align*}
	O_A \ket{\psi_1} = \omega_p^{-1} \ket{\psi_1}, &&
	O_B \ket{\psi_1} = \omega_p \ket{\psi_1}.
\end{align*}
Since $S$ can also induce $P_{\hat{A}}$, and $P_{\hat{A}}$ with Alice and Bob's roles flipped,
by \cref{prop:hata}, 
we know
\begin{align*}
	&U_AO_AU_A\ct \ket{\psi} = O_A^r \ket{\psi}, \\
	&U_BO_BU_B\ct \ket{\psi} = O_B^r \ket{\psi}.
\end{align*}
By substitution and \cref{lm:sub}, we know
\begin{align*}
	&O_A (U_A\ct)^j \ket{\psi} = (U_A \ct)^j O_A^{r^j} \ket{\psi}, \\
	&O_B (U_B\ct)^j \ket{\psi} = (U_B \ct)^j O_B^{r^j} \ket{\psi},
\end{align*}
for $j \in [p-1]$.
Then, 
\begin{align*}
	&O_A (U_A\ct)^j \ket{\psi_1}  = (U_A \ct)^j O_A^{r^j} \ket{\psi_1} = \omega_p^{-r^j} (U_A \ct)^j \ket{\psi_1}, \\
	&O_B (U_B\ct)^j \ket{\psi_1}  = (U_B \ct)^j O_B^{r^j} \ket{\psi_1} = \omega_p^{r^j} (U_B \ct)^j \ket{\psi_1},
\end{align*}
where we use the fact that $\ket{\psi_1}$ can be expressed using Alice's projectors and Bob's projectors as shown in \Cref{eq:psi1,eq:psi1_q}.
Hence, $\set{ (U_A\ct U_B\ct)^j \ket{\psi_1} | j \in [p-1]}$ is an orthogonal set
and 
\begin{align*}
	\norm{  \sum_{j \in [p-1]} (U_A\ct U_B\ct)^{j} \ket{\psi_1}} = 1. 
\end{align*}
The last step is to check that $\bra{\psi} \sum_{j \in [p-1]} (U_A\ct U_B\ct)^{j} \ket{\psi_1} = 1$.
Using the fact that 
$U_AU_B\ket{\psi} = P_{u_1} Q_{u_1} P_{u_2}Q_{u_2} \ket{\psi} = \ket{\psi}$, we can see that 
\begin{align*}
	\bra{\psi} \sum_{j \in [p-1]} (U_A\ct U_B\ct)^{j} \ket{\psi_1} 
	= (p-1) \braket{\psi}{\psi_1}.
\end{align*}	
Hence, the problem is reduced to calculate $\braket{\psi}{\psi_1}$, which is
\begin{align*}
	&\frac{1}{2} \bra{\psi}(P_{\nr+\mr+1}^{(0)} +P_{\nr+\mr+1}^{(1)} \\
                       &-iP_{\nr+\mr+2}P_{\nr+\mr+1}) \ket{\psi} \\
                    = &\frac{1}{p-1} - \frac{i}{2} \bra{\psi} Q_{\nr+\mr+2}P_{\nr+\mr+1}\ket{\psi} \\
                    = &\frac{1}{p-1},
\end{align*}	
where $\bra{\psi} Q_{\nr+\mr+2}P_{\nr+\mr+1}\ket{\psi} = 0$ comes from the correlation.
Then the proposition follows.
\end{proof}
\begin{theorem}
	\label{thm:self-test}
	Let $S$ be an inducing strategy of $P_{p,r}$ with a shared state $\ket{\psi}$.
	Then there exist an isometry   
	$\Phi_A \x \Phi_B$ and a state $\ket{junk}$
	such that 
	\begin{align*}
		&\Phi_A \x \Phi_B (\ket{\psi}) = \ket{junk} \x \ket{\tpsi}\\	
		&\Phi_A \x \Phi_B (O_A \ket{\psi})
		= \ket{junk}  \x [(O_1O_2) \x \1] \ket{\tpsi} \\
		&\Phi_A \x \Phi_B (O_B\ket{\psi}) 
		= \ket{junk}  \x[\1 \x (O_1O_2)] \ket{\tpsi} \\
		&\Phi_A \x \Phi_B (U_A  \ket{\psi}) 
		= \ket{junk}  \x [(U_1U_2) \x \1] \ket{\tpsi} \\
		&\Phi_A \x \Phi_B (U_B\ket{\psi})
		= \ket{junk}  \x [\1 \x (U_1U_2) ] \ket{\tpsi} ,
	\end{align*}
	where $\ket{\tpsi}$ is defined in \Cref{eq:tpsi} and
	$O_1, O_2, U_1, U_2$ are defined in \Cref{eq:u1,eq:u2,eq:o1,eq:o2}
	respectively.
\end{theorem}
The isometry $\Phi_A \x \Phi_B$ is given in the figure below. It is designed based on
the swap isometry proposed in \cite{yang2013}.
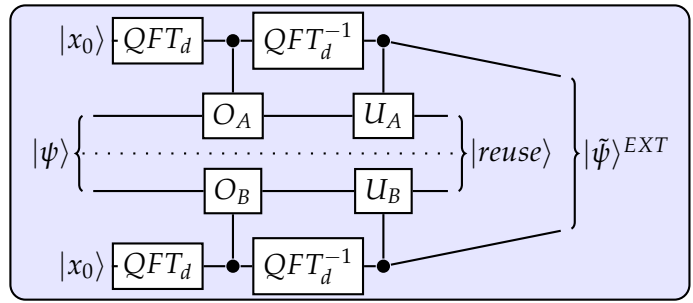
\begin{figure}[H]
\center
        \begin{tikzpicture}[thick]
        %
        \tikzstyle{operator} = [draw,fill=white,minimum size=1.5em] 
        \tikzstyle{phase} = [fill,shape=circle,minimum size=5pt,inner sep=0pt]
        \tikzstyle{surround} = [fill=blue!10,thick,draw=black,rounded corners=2mm]
        %
        \draw[decorate,decoration={brace,mirror},thick] (0,-1) to
    	node[midway,left] (bracket1) {$\ket{\psi}$}
    	(0,-2);
        \node at (0,0) (q1) {$\ket{x_0}$};
        \node at (0,-1) (q2) {};
        \node at (0,-2) (q3) {};
        \node at (0,-3) (q4) {$\ket{x_0}$};
        %
        \node[operator] (op11) at (1,0) {$QFT_p$} edge [-] (q1);
        \node[operator] (op14) at (1,-3) {$QFT_p$} edge [-] (q4);
        %
        \node[phase] (phase11) at (2,0) {} edge [-] (op11);
	\node[operator] (op22) at (2,-1) {$O_A$} edge [-] (q2);
	\node[operator] (op23) at (2, -2) {$O_B$} edge[-] (q3);
        \node[phase] (phase14) at (2,-3) {} edge [-] (op14);
        \draw[-] (phase11) -- (op22);
        \draw[-] (phase14) -- (op23);
        %
        \node[operator] (op31) at (3,0) {$QFT_p^{-1}$} edge [-] (phase11);
        \node[operator] (op34) at (3,-3) {$QFT_p^{-1}$} edge [-] (phase14);
        %
        \node[phase] (phase21) at (4,0) {} edge [-] (op31);
	\node[operator] (op42) at (4,-1) {$U_A$} edge [-] (op22);
	\node[operator] (op43) at (4, -2) {$U_B$} edge[-] (op23);
        \node[phase] (phase24) at (4,-3) {} edge [-] (op34);
        \draw[-] (phase21) -- (op42);
        \draw[-] (phase24) -- (op43);
        %
        \node (end2) at (5,-1) {} edge [-] (op42);
        \node (end3) at (5,-2) {} edge [-] (op43);
        %
        \draw[decorate,decoration={brace},thick] (5,-1) to
    	node[midway,right] (bracket) {$\ket{junk}$}
    	(5,-2);
        %
        \node (end1) at (6.5,-0.5) {} edge [-] (phase21);
        \node (end4) at (6.5,-2.5) {} edge [-] (phase24);
        \draw[loosely dotted] (0,-1.5) -- (5,-1.5);
        \draw[decorate,decoration={brace},thick,] (6.5,-0.5) to
    	node[midway,right] (bracket2) {$\ket{\tpsi}$}
    	(6.5,-2.5);
        %
        \begin{pgfonlayer}{background} 
        \node[surround] (background) [fit = (q1) (op14) (bracket1)(bracket2)] {};
        \end{pgfonlayer}
        \end{tikzpicture}
	\caption{The isometries $\Phi_{A} \x \Phi_{B}$.}
\end{figure}
The isometry $\Phi_{A} \x \Phi_{B}$ has the following steps:
\begin{enumerate}
	\item Append control register $\ket{x_0}_{A'}$ on Alice's side and $\ket{x_0}_{B'}$ on Bob's side, where $\ket{x_0}$
	is orthogonal to $\C^{p-1}$;
	\item Apply Quantum Fourier Transform ($QFT_p$) to Alice and Bob's control registers;
	\item Apply Controlled-$O_{A/B}$ operations (i.e. if the control register is in state $\ket{x_k}_{A'/ B'}$, apply
	$O_{A/B}^k$.);
	\item Apply inverse Quantum Fourier Transform ($QFT_p^{-1}$) to the control registers;
	\item Apply Controlled-$U_{A/B}$ operations (i.e. If Alice's control register is in state $\ket{x_j}$, she applies
	$U_A^{\log_r (p-j)}$. If Bob's control register is in state $\ket{x_j}$, he applies $(U_B)^{\log_r j}$).
\end{enumerate}
\begin{proof}
	\Cref{prop:decomp_psi} implies that 
	\begin{align*}
	 \Phi_{A} \x \Phi_{B} (\ket{\psi}) = \Phi_{A} \x \Phi_{B} (\sum_{j=1}^{(p-1)} (U_A\ct U_B\ct)^{\log_r j} \ket{\psi_1} ).
	\end{align*}
	Then the state $\ket{\psi}$ is evolved by the isometry as follows
	\begin{align*}
		& \sum_{j=1}^{p-1} (U_A\ct U_B\ct)^{\log_r j} \ket{\psi_1} \ket{x_0}_{A'}\ket{x_0}_{B'}\\
		\xrightarrow[]{QFT_p}& \frac{1}{p}\sum_{k_1,k_2 \in [p]} \sum_{j=1}^{p-1} (U_A\ct U_B\ct)^{\log_r j}  \\
		 &\times \ket{\psi_1}\ket{x_{k_1}}_{A'}\ket{x_{k_2}}_{B'}\\
		\xrightarrow[]{\text{Controlled-}O_{A/B}}& \frac{1}{p}\sum_{k_1,k_2 \in[p]} \sum_{j=1}^{p-1} O_A^{k_1}(U_A\ct)^{\log_r j}  \\
		& \times O_B^{k_2}(U_B\ct)^{\log_r j} \ket{\psi_1} \ket{x_{k_1}}_{A'}\ket{x_{k_2}}_{B'}\\
		=&\frac{1}{p} \sum_{k_1,k_2 \in [p]} \sum_{j=1}^{p-1} (U_A\ct U_B\ct)^{\log_r j}  \\
		&\times \omega_p^{(k_2-k_1)j}\ket{\psi_1} \ket{x_{k_1}}_{A'}\ket{x_{k_2}}_{B'}\\
		\xrightarrow[]{QFT_p^{-1}} &\frac{1}{p^2}\sum_{j=1}^{p-1} \sum_{l_1,l_2 \in [p]} (U_A\ct U_B\ct)^{\log_r j} \\
		&\times \left(\sum_{k_1 \in [p]}\omega_p^{k_1(p-j-l_1)}\right) \\
		& \times  \left(\sum_{k_2\in[p]}\omega_p^{k_2(j-l_2)}\right)
		\ket{\psi_1} \ket{x_{l_1}}_{A'}\ket{x_{l_2}}_{B'}\\
		= &\sum_{j=1}^{p-1}(U_A\ct U_B\ct)^{\log_r j} \ket{\psi_1} \ket{x_{p-j}}_{A'}\ket{x_j}_{B'} \\
		\xrightarrow[]{\text{Controlled-}U_{A/B}}& \sum_{j=1}^{p-1} U_A^{\log_r j} (U_A\ct)^{\log_r j} U_B^{\log_r j} (U_B\ct)^{\log_r j}\\
		&\times \ket{\psi_1} \ket{x_{p-j}}_{A'}\ket{x_j}_{B'}\\
		=&\ket{\psi_1} \x \sum_{j=1}^{p-1} \ket{x_{p-j}}_{A'}\ket{x_j}_{B'},
	\end{align*}
	where we use the fact that for $l \in [p]$,
	\begin{align*}
		\sum_{k \in [p]} \omega_p^{k(j-l)} =\begin{cases} 
		p&\text{ if } l = j \\
		0 &\text{ otherwise.}
		\end{cases}
	\end{align*}
	The derivations of the other four equations follow very similar argument, so instead of giving
	the full proof, we give the key steps for them.
	For $O_A$ and $O_B$, we use the fact that 
	\begin{align*}
		&O_A  \sum_{j=1}^{p-1} (U_A\ct U_B\ct)^{\log_r j} \ket{\psi_1} 
		= \sum_{j=1}^{p-1}  \omega_p^{-j} (U_A\ct U_B\ct)^{\log_r j}\ket{\psi_1},\\
		&O_B  \sum_{j=1}^{p-1} (U_A\ct U_B\ct)^{\log_r j} \ket{\psi_1} 
		= \sum_{j=1}^{p-1}  \omega_p^{j} (U_A\ct U_B\ct)^{\log_r j}\ket{\psi_1}.
	\end{align*}
	For $U_A$ and $U_B$, we use the fact that
	\begin{align*}
		&O_A (U_A\ct)^{\log_r j -1} \ket{\psi_1} = \omega_p^{-jr^{-1}} (U_A\ct)^{\log_r j -1} \ket{\psi_1}, \\
		&O_B (U_B\ct)^{\log_r j -1} \ket{\psi_1} = \omega_p^{jr^{-1}} (U_B\ct)^{\log_r j -1} \ket{\psi_1},
	\end{align*}
	for all $1 \leq j \leq p-1$.
\end{proof}
\Cref{thm:self-test} implies that,
for any odd prime number $p$ whose primitive root is $r$, 
there exists a correlation of size $\Theta(r^2)$ that can self-test a maximally
entangled state of local dimension $p-1$.
Since there exists $r \in \set{2,3,5}$ such that there 
are infinitely many prime numbers
whose smallest primitive root is $r$ \cite{murty1988},
we can apply \cref{thm:self-test} to the set $D$
of all such odd prime
numbers and obtain \Cref{thm:inf}.

\section{Conclusion and discussions}
\label{sec:conclude}
We have shown that there exists a family of constant-sized correlations such that
each correlation of this family can self-test a maximally entangled state of a different dimension.

Curious readers may wonder if we can self-test $\ket{\tpsi} \x \ket{EPR}^{\x 2}$ since the inducing
strategy uses it. Indeed, we can self-test it,
but the proof of the self-test requires more details of the representation of $\Gamma(\hat{A})$ and modifying $\hat{A}x=0$ to 
introduce equations of the Magic square game. Since the techniques to self-test $\ket{EPR}^{\x 2}$ is standard
in the literature, we refer to \cite[Section 6]{fu2019}, which is the previous arXiv version of this paper, for details of self-testing $\ket{\tpsi} \x \ket{EPR}^{\x 2}$.

The other question to ask is if we can self-test $O_1$ and $O_2$ individually.
We don't have an answer for this question. 
The progress of answering this question is summarized here.
Following the calculation in the proof of \cref{prop:pmu_ext}, we can see that
\begin{align*}
	&(P_1P_2) P_2 \ket{\psi_1} = \omega_p P_2 \ket{\psi_1},\\
	&(Q_1Q_2) Q_2 \ket{\psi_1} = \omega_p^{-1} Q_2 \ket{\psi_1}.
\end{align*}
Then, to self-test $O_1$ and $O_2$ on Alice's side, we need to modify the controlled-$U_{A/B}$ step
so that when the control-register is in the state $\ket{x_j}$, both Alice and Bob apply $U_{A/B}^{\log_r j}$.
If we denote the modified isometry by $\Phi_A' \x \Phi_B'$,
\begin{align*}
	&\Phi_A' \x \Phi_B'(P_{o_x} \ket{\psi}) = \ket{junk} \x (O_x \x \1) \ket{\tpsi},
\end{align*}
with $x= 1, 2$.
What is interesting is that if we want to self-test $O_1$ and $O_2$ on Bob's side, we need to 
modify the controlled-$U_{A/B}$ step again,
so that when the control-register is in the state $\ket{x_j}$, both Alice and Bob apply $U_{A/B}^{\log_r (p-j)}$.
If we denote the modified isometry by $\Phi_A'' \x \Phi_B''$,
\begin{align*}
	&\Phi_A'' \x \Phi_B''(Q_{o_y} \ket{\psi}) = \ket{junk} \x (\1 \x O_y) \ket{\tpsi},
\end{align*}
with $y= 1, 2$.
Since $\Phi_A' \x \Phi_B'$ and 
$\Phi_A'' \x \Phi_B''$ are different from $\Phi_A \x \Phi_B$, we cannot conclude that $P_{p,r}$ is a self-test of $\ket{\tpsi}$ and the binary observables $O_1$ and $O_2$.
On the other hand, it has been shown projectors of arbitrary rank can be self-tested \cite{mancinska2021}.
If we can self-test $O_1$, $O_2$ and even
$U_1$ and $U_2$, we can get a second self-test of projectors of arbitrary rank.

The last question about $P_{p,r}$ is about its robustness. In the arXiv version \cite{fu2019}, we have shown that 
if a quantum strategy $S$ can induce a correlation $P'$ such that 
\begin{align*}
	\max_{x,y,a,b}  \quad \abs{P'(a,b|x,y) - P_{p,r}(a,b | x,y)} \leq \epsilon,
\end{align*}
then,
\begin{align*}
	\norm{\Phi_A \x \Phi_B(\ket{\psi}) - \ket{junk} \x \ket{\tpsi}} = O(r^p \ep^{1/8}).
\end{align*}
The derivation of the robustness is very long and technical, so we choose the journal version to focus on
illustrating the basic ideas. Improving the robustness bound is another open problem about $P_{p,r}$.

\section*{Acknowledgment}
The author is deeply grateful to his PhD advisor, Carl Miller, for the enlightening discussions with him.
Carl's original ideas about combining a linear system game with a Bell inequality and about characterizing prime numbers
by their primitive roots are invaluable to this work. Carl also helped the author with the design
of $\hat{P}_{-\pi/p}$.

\bibliographystyle{plainnat}
\bibliography{quantum_correlation}
\end{document}